\documentclass[aps,prl,twocolumn,showpacs,preprintnumbers,amsmath,amssymb]{revtex4}

\usepackage{amssymb}
\usepackage{url}
\usepackage{theorem}
\newcommand{\mcM}{{\mycal M}}
\newcommand{\mcK}{{\mycal K}}

\newcommand{\Ricc}{\mathrm{Ric}\,}

\newcommand{\bea}{\begin{eqnarray}}
\newcommand{\beaa}{\begin{eqnarray*}}
\newcommand{\bean}{\begin{eqnarray}\nonumber}
\newcommand{\bel}[1]{\begin{equation}\label{#1}}
\newcommand{\beal}[1]{\begin{eqnarray}\label{#1}}
\newcommand{\beadl}[1]{\begin{deqarr}\label{#1}}
\newcommand{\eeadl}[1]{\arrlabel{#1}\end{deqarr}}
\newcommand{\eeal}[1]{\label{#1}\end{eqnarray}}
\newcommand{\eead}[1]{\end{deqarr}}
\newcommand{\eea}{\end{eqnarray}}
\newcommand{\eeaa}{\end{eqnarray*}}

\newcommand{\be}{\begin{equation}}
\newcommand{\ee}{\end{equation}}

\newcommand{\tr}{\mbox{\rm tr}}

\DeclareFontFamily{OT1}{rsfs}{}
\DeclareFontShape{OT1}{rsfs}{m}{n}{ <-7> rsfs5 <7-10> rsfs7 <10->
rsfs10}{} \DeclareMathAlphabet{\mycal}{OT1}{rsfs}{m}{n}

\usepackage{amsmath} 

\newtheorem{Theorem}{\sc Theorem\rm}

\def \Reel{\mathbb{R}}

\def \R {\Reel}

\def \Nat{\mathbb{N}}

\def \N {\Nat}

\newcommand{\Sp}[0]{\mbox{$\mathbb{S} $}}

\newcounter{mnotecount}[section]

\newcommand{\rmnote}[1]{}

\begin{document}
\title{Gluing Initial Data Sets for General Relativity}

\author{Piotr T. Chru\'sciel}
\thanks{Partially supported by a Polish Research Committee 
        grant 2 P03B 073 24}
\email{Piotr.Chrusciel@lmpt.univ-tours.fr}
\homepage{www.phys.univ-tours.fr/~piotr}
\affiliation{D\'epartm\'ent de Math\'ematiques, Facult\'e des Sciences,
Universit\'e de Tours, Parc de Grandmont, F37200, Tours, France}

\author{James Isenberg}
\thanks{Partially supported by the NSF under Grant PHY-0099373}  
\email{jim@newton.uoregon.edu} 
\homepage{www.physics.uoregon.edu/~jim} 
\affiliation{Department of Mathematics, University of Oregon, Eugene, 
Oregon 97403-5203, USA}

\author{Daniel Pollack}
\thanks{Partially supported by the NSF under Grant DMS-0305048 
        and the UW Royalty Research Fund}
\email{pollack@math.washington.edu}
\homepage{www.math.washington.edu/~pollack}
\affiliation{Department of Mathematics, University of Washington,
Box 354350, Seattle, Washington 98195-4350, USA}

\date{18 August, 2004}

\begin{abstract}
We establish an optimal gluing construction for general relativistic initial 
data sets.  The construction is optimal in two distinct ways. First, 
it applies to {\it generic} initial data sets and the required 
(generically satisfied) hypotheses are geometrically and physically natural.
Secondly, 
the construction is completely {\it local} in the sense that the 
initial data is left
unaltered on the complement of arbitrarily small neighborhoods of the 
points about which the gluing takes place.  Using this construction we 
establish the existence of cosmological, maximal globally hyperbolic, vacuum 
space-times with no constant mean curvature spacelike Cauchy surfaces.
\end{abstract}

\pacs{04.20.Ex}
\maketitle

\section{Introduction}

As with Maxwell's theory, a set of initial data for Einstein's theory of 
gravity must satisfy a set of constraint equations. Unlike the Maxwell 
constraints $\nabla\cdot E=\rho$ and $\nabla \cdot B=0$, the Einstein 
constraint equations are nonlinear and fairly complicated. Hence, although 
much has been learned about solutions of the Einstein constraint equations 
during the past thirty years using the conformal method and related 
techniques, many questions concerning them have remained formidable. 
A number of these questions can now be answered using a powerful tool from 
geometric analysis: Gluing. Analytic gluing techniques have played a prominent
role in many areas of differential geometry over the past twenty years. 
Some particularly notable applications include: 
the study of smooth topology of four manifolds 
(via Donaldson and later Seiberg-Witten theory), 
pseudo-holomorphic curves in symplectic 
geometry, the existence of half-conformally flat structures on four manifolds,
manifolds with exceptional holonomy, singular Yamabe metrics  
and  the study of minimal and constant mean curvature surfaces
in Euclidean three-space, and many others.

It is only recently that these techniques have been applied to general 
relativity, and the impact of this work has been significant.
In the first two applications gluing techniques were used in very 
different ways  toward quite different aims.  In \cite{Corvino}, Corvino 
applied a new gluing method to asymptotically flat, time symmetric initial 
data.  The core feature of \cite{Corvino} is a local deformation result 
for the scalar curvature operator.  By exploiting the underdetermined nature 
of this operator (or more precisely, the fact that its adjoint is
overdetermined) Corvino is able to show that one can solve for prescribed, 
small  {\em compactly supported} deformations of the scalar curvature. 
This deformation result is used in  \cite{Corvino} to prove the existence of 
asymptotically flat, scalar flat metrics on $\R^n$ $(n\geq 3)$ which are
Schwarzschild outside of a compact set.  
The evolution of this initial data produces nontrivial space-times 
which are identically Schwarzschild near spatial infinity. 
In the more general setting of constant mean curvature initial data 
sets a gluing construction was developed \cite{IMP}, in the context of 
the well known conformal method of Lichnerowicz, Choquet-Bruhat and York
which reduces the constraints equations to a determined elliptic system.
The construction of \cite{IMP}, and subsequently \cite{IMP2},
allowed one to demonstrate how space-times can be joined by means of a
geometric connected sum, or how a wormhole can be added between two 
points in a given space-time (on the level of the initial data).
This was flexible enough to address a number of 
issues concerning the relation of the spatial topology to the geometry of 
solutions of the constraints and 
the constructibility of multi black hole solutions (see also \cite{ChMazzeo}). 

We have subsequently developed this technique so that it can be 
applied to a much wider class of solutions; 
indeed, we have now obtained the sharpest possible gluing theorem for the 
Einstein constraint equations.  Using it, we can do all of the following:
1) Show that for a generic solution of the constraint equations and any 
pair of points in this solution, one can add a wormhole 
connecting these points to the solution with  {\em no} change in the 
data away from a neighborhood of each of the points. 2) Show that for 
almost any pair of initial data sets (including, say, a pair of black 
hole data sets, or a cosmological data set paired with a set of black 
hole data) one can construct a new set which joins them.  3) Prove that 
there exist spatially compact maximal globally hyperbolic space-times 
which satisfy the vacuum Einstein equations and contain no closed 
constant mean curvature hypersurface. It is likely that 
these new gluing techniques will continue to be very useful for the practical 
construction of physically interesting initial data sets.
 
\section{Main results}

We recall that a set of initial 
data $(M^n, \gamma, K, \Psi)$ for Einstein's theory consists 
of an n-manifold $M^n$, a Riemannian metric $\gamma$ on $M^n$, 
a symmetric tensor field $K$ on $M^n$, and possibly a set of 
non-gravitational fields $\Psi$ (e.g., $E$ and $B$ for Einstein-Maxwell). 
The constraint equations require that $(M^n, \gamma, K, \Psi)$ 
satisfy
 \beal{ce20n} & {16 \pi \rho = R(\gamma) -(2\Lambda+
|K|_\gamma^2-(\tr_\gamma K)^2)}\;, &\label{ce-nv+cc1}
 \\ & 16 \pi J^j = 2 D_i(K^{ij}-\tr_\gamma K \gamma^{ij})\;.
&\label{ce-nv+cc2}\eea
where $D$ is the covariant derivative corresponding to $\gamma$, 
$R$ is its scalar curvature, $\rho=\rho(\gamma, \Psi)$ is the energy density 
function of the non-gravitational fields, and $J=J(\gamma, \Psi)$ is the 
corresponding momentum flow vector field
\footnote{For Maxwell theory, $\rho=\frac{1}{2}(E^2+B^2)$, and $J=E\times B$}.
Now let $(M^n, \gamma, K, \Psi)$ be a solution of 
(\ref{ce-nv+cc1})-(\ref{ce-nv+cc2}), and let $p_1$ and $p_2$ 
be a pair of points contained in $M^n$.  
The basic idea of gluing is simple: let $\tilde M$ be the manifold
obtained by removing from $M$ geodesic balls of radius $\epsilon$ 
around $p_1$ and $p_2$, and gluing in a neck $\Sp^{n-1}\times I$,
One  then tries to construct initial data 
$(\tilde \gamma(\epsilon), \tilde K(\epsilon), \tilde\Psi(\epsilon))$ 
on $\tilde M$ which  coincide with the original data
away from a small neighborhood of the neck. If the points lie in distinct
connected components of $M$, then the manifold $\tilde M$ is the connected
sum of those components. If the points lie in the same component, then 
$\tilde M$ consists of $M$ together with a ``handle'' or wormhole connecting
neighborhoods of the two points.

We cannot expect to be able to glue every pair of solutions of the 
constraints. For example, if we could \textit{locally}
glue a set of data for Minkowski 
space to a solution of the constraints on a manifold which does not 
admit a flat metric, then the resulting data  would have zero ADM mass, 
and yet would not be data for Minkowski space, thereby violating the 
positive energy theorem \cite{SchoenYau79b}. 
So there are necessarily conditions 
a data set must satisfy if it is to admit a gluing construction as above.
 
As noted above, in earlier work, it was required that a solution 
have constant mean curvature (CMC) \cite{IMP}, or at least have 
a CMC region surrounding each of the chosen gluing points 
\cite{IMP2}. Further,  global  ``nondegeneracy" 
conditions needed to be satisfied as well. Consequently, gluing could not be 
applied generically. Our results here are much less restrictive. 
The condition which must be met is local, in the sense that it 
only involves the data in regions close to the gluing points. 
Further, the condition is satisfied at all points in generic solutions. 
 
To define the gluing condition in the vacuum case, we fix a 
solution $(M^n, \gamma, K)$ of the vacuum constraint equations  
(\ref{ce-nv+cc1})-(\ref{ce-nv+cc2}) (with $\rho=0$, and $J=0$), 
and consider the $L^2$ adjoint $\mathcal{P}^*_{(\gamma, K)}$ 
of the linearization of the constraint equations at this solution. 
Viewed as an operator acting on a scalar function $N$ and a 
vector field $Y$, $\mathcal{P}^*_{(\gamma, K)}$ 
takes the explicit form

\begin{widetext}
\begin{equation}
\label{4} \mathcal{P}^*_{(\gamma, K)}(N,Y)=\left(
\begin{array}{l}
2(\nabla_{(i}Y_{j)}-\nabla^lY_l g_{ij}-K_{ij}N+\tr K\; N g_{ij})\\
 \\
\nabla^lY_l K_{ij}-2K^l{}_{(i}\nabla_{j)}Y_l+
K^q{}_l\nabla_qY^lg_{ij} -\Delta N g_{ij}+\nabla_i\nabla_j N \\ \;
+(\nabla^{p}K_{lp}g_{ij}-\nabla_lK_{ij})Y^l
-N \Ricc(g)_{ij} +2NK^l{}_iK_{jl}-2N (\tr \;K) K_{ij}
\end{array}
\right)\;.
\end{equation}
\end{widetext}
 Now let $\Omega$ be an open subset of $M^n$. By definition, the set of 
``KIDs" on $\Omega$, denoted $\mcK(\Omega)$, 
is the set of all solutions of the equation 
 \begin{equation} 
 \mathcal{P}^*_{(\gamma, K)|_{\Omega}} (N,Y)=0.
 \label{NoKIDs}
 \end{equation} 
Such a solution $(N,Y)$, if nontrivial, generates a space-time 
Killing vector field \footnote{Hence the name ``KIDs", 
which abbreviates ``Killing initial data set"} 
in the domain of dependence of
$(\Omega, \gamma|_{\Omega}, K|_{\Omega})$~\cite{MonLSI}.
In terms of KIDs, our only ``nondegeneracy condition'' 
is simply that there exist 
neighborhoods $\Omega_1 \ni p_1$ and $\Omega_2 \ni p_2$ of the 
gluing points such that the KIDs (with no boundary conditions imposed)
are all trivial on $\Omega_1$ and $\Omega_2$.
 
 We can now formally state our main result, for vacuum initial data:
 \begin{Theorem}
\label{Tlgluingv} Let $( M, \gamma, K)$ be a
smooth vacuum initial data set, and consider two open sets
$\Omega_a\subset  M$  such that 
\begin{equation}
\mbox{ the set of KIDs, $\mcK(\Omega_a)$, is
trivial.} 
\label{nokids}
\end{equation}
Then for all  $p_a\in \Omega_a$, $\epsilon >0$ and
$k\in \N$ there exists a smooth vacuum initial data set
$(\tilde \gamma(\epsilon),\tilde K(\epsilon))$ on $\tilde M$ such that
$(\tilde \gamma(\epsilon),\tilde K(\epsilon))$ is $\epsilon$-close to  
$(\gamma, K)$ in a $C^k\times C^k$ topology away from
$B(p_1,\epsilon)\cup B(p_2,\epsilon)$. Moreover
$(\tilde \gamma(\epsilon), \tilde K(\epsilon))$ coincides with 
$(\gamma, K)$ away from $\Omega_1\cup \Omega_2$.
\end{Theorem}
While the tie between nontrivial KIDs and the presence of local Killing 
fields suggests that the absence of nontrivial KIDs is generic, such a 
result requires proof. Theorems to this effect are proven in 
\cite{ChBeignokids}.

Besides the significant relaxing of the gluing conditions which this new 
result provides, the fact that the glued data is identical to the 
original data away from the points $p_1$ and $p_2$ in 
Theorem \ref{Tlgluingv} provides an important
improvement over the earlier gluing results \cite{IMP} and \cite{IMP2}.
These earlier results only guaranteed that the glued data set is arbitrarily 
close (relative to an appropriate function space) to what it was
originally.

What happens for solutions of the constraints with non-gravitational 
``matter" fields present? If the fields are entirely described by 
the choice $\rho$ and $J$, then the conditions needed for gluing 
are relatively mild: It is sufficient that $\rho(x)>|J(x)|$ for all 
$x$ in a pair of neighborhoods of the points $p_1$ and $p_2$ at 
which gluing is to be done. Since the energy condition $\rho(x)\geq |J(x)|$ 
is generally imposed for physical reasons, requiring that the inequality 
be strict is a mild additional restriction. For non-gravitational  fields 
with the introduction of additional constraint equations (e.g., the 
Einstein-Maxwell theory which adds $\nabla \cdot E=0$ and 
$\nabla \cdot B=0$), it is likely that the required condition for a
local gluing construction
is a natural generalization of the ``no KIDs" condition.

\section{The construction}

The detailed proofs of Theorem \ref{Tlgluingv} and of the 
analogous Einstein-matter theorem (with no extra constraints) 
are described in \cite{IDE}. Here, to illustrate some of the 
ideas involved, we provide a brief sketch of the vacuum case.
We choose balls $B(p_1,r_1)\subset \Omega_1$ and $B(p_2,r_2)\subset \Omega_2$
within which to do the gluing.
In~\cite{ChBeignokids} it is
shown that, under the nondegeneracy assumption (\ref{nokids}), we can
$\epsilon$-perturb the data on $\Omega_1$ and $\Omega_2$, without
changing them away from those regions, so that the constraint
equations still hold, and so that there are no space-time
isometries in any open set contained within $B(p_a,r)$, for
sufficiently small $r$.
The next step is to use a theorem of
Bartnik~\cite{bartnik:variational} to deform the balls $B(p_a,r)$,
in the space-time evolution of this new data, 
so that the trace of $K$ is constant on $B(p_a,r)$,
reducing $r$ if necessary. The non-existence of space-time
isometries is preserved under this deformation. This deformation
is done so that we are in the setting in which a generalization of
the gluing theorem of~\cite{IMP} to compact manifolds with
boundary (and to include matter fields) may be applied. This
constitutes the third step in the construction and is essentially done by
repeating the arguments of~\cite{IMP,IMaxP} in this new setting.
We thus obtain a one parameter family of initial data which satisfies
the constraint equations, and which contains a neck connecting 
the spheres $S(p_a,r)$. This family of data has
the property that the initial data approach the original ones in a
neighborhood of the $S(p_a,r)$'s when the parameter $\epsilon$ 
tends to zero. 
However, the transverse derivatives of those data do not match
those of the original ones at the boundary spheres. This problem is cured,
for $\epsilon$ small enough, by a theorem in~\cite{ChDelay}, which
holds precisely under the ``no local space-time isometries"
condition (\ref{nokids}). 
This provides the desired gluing, localized within the sets $\Omega_a$.

\section{Space-times with no CMC slices}

One of the original motivations for attempting to apply gluing 
constructions to initial data has been to show that 
there are spatially compact, maximally extended, globally 
hyperbolic solutions of the vacuum Einstein equations with no
constant mean curvature Cauchy surfaces. This result is interesting, 
since the traditional view of both mathematical and numerical relativists 
has been that the most useful and reliable choice of time for a globally 
hyperbolic space-time is one based on a foliation by CMC slices. Such a 
foliation, if it exists in a given solution, has the virtue that it is 
unique \cite{BF78};
CMC slices also appear to avoid singularities in numerical simulations. 

In \cite{bartnik:cosmological}, 
Bartnik shows that there exist maximally extended, globally 
hyperbolic solutions of the Einstein equations \textit{with dust} 
which admit no 
CMC slices. Later, Eardley and Witt proposed a scheme for showing that 
similar vacuum solutions exist \cite{EW92}, but their proof was incomplete. 
Using our gluing result in Theorem \ref{Tlgluingv}, we can now complete the 
argument.

We consider a set of vacuum initial data 
$(T^3\#T^3, \gamma, K)$, where $T^3\#T^3$ is the connected sum of a pair 
of three tori. We assume that, relative to some chosen two sphere $S$ 
on the connecting cylinder of $T^3\#T^3$, there is a ``reflection map" 
$\mu:T^3\#T^3 \rightarrow T^3\#T^3$ with the following properties: (i) 
$\mu$ is a diffeomorphism; (ii) $\mu(S)=S$; (iii) $\mu^*(\gamma)=\gamma$; 
and (iv) $\mu^*(K)=-K$. Note that as a consequence of these properties, 
$K|_S=0$.

It follows from \cite{CBG69} that there is a unique, maximally extended,
globally hyperbolic, development $(T^3\#T^3 \times \R, g)$ of the data 
$(T^3\#T^3, \gamma, K)$, with $g$ satisfying the Einstein vacuum field 
equations on $T^3\#T^3 \times \R$. 
Further, as a consequence of this uniqueness, 
the map $\mu$ described above extends to a diffeomorphism from 
$(T^3\#T^3 \times \R, g)$ to itself  with the property that if $\Sigma^3$ 
is a Cauchy surface in $(T^3\#T^3 \times \R, g)$ with induced data 
$(\gamma_{\Sigma^3}, K_{\Sigma^3})$, then $\mu(\Sigma^3)$ 
is a Cauchy surface as well, with its data 
$(\gamma_{\mu(\Sigma^3)}, K_{\mu(\Sigma^3)})$ satisfying
\begin{equation}
\mu^*(\gamma_{\mu(\Sigma^3)})=\gamma
\qquad\mbox{and}\qquad
\mu^*(K_{\mu(\Sigma^3)})=-K.
\label{inducedmetric+K}
\end{equation}
Say there exists a CMC Cauchy surface $\Sigma_{\tau}$ in the space-time 
$(T^3\#T^3 \times \R, g)$ with mean curvature $tr(K_{\Sigma_{\tau}})=\tau$ 
constant. Applying the map $\mu$, we obtain another Cauchy surface, which 
must also have constant mean curvature, but with 
$tr(K_{\mu(\Sigma_{\tau})})=-\tau$. It now follows from barrier 
arguments that there must be a \textit{maximal} Cauchy surface 
in the space-time (with $\tau=0$). However, if there is such a Cauchy surface, 
then it follows from the constraint equations 
that the scalar curvature on this maximal surface must be non-negative. 
This is known to be incompatible with the topology $T^3\#T^3$ 
\cite{SchoenYau79a}. Thus we have a contradiction, 
from which it follows that the space-time development of initial data with 
the reflection properties described above cannot contain a CMC Cauchy surface. 

We now use Theorem~\ref{Tlgluingv} to show that we can produce such data. 
To start, we use the conformal method to find a CMC solution 
$(T^3, \hat{\gamma}, \hat{K})$ of the constraints on the torus which has 
no conformal Killing fields, has nonvanishing mean curvature, and has the 
traceless part of $\hat{K}$ nonvanishing. (It follows from 
\cite{ChBeignokids, bergerebin, Iconstraints} that such data sets exist.) 
We easily verify that consequently this data has no global KIDs, 
$\mcK(T^3)=0$.

Let $(\mcM,g)$ be the maximal, globally hyperbolic, development of
this data. We deform the initial
data hypersurface $T^3$ in $\mcM$ to create a small  neighborhood of a
point $p$ in which the trace of the new induced $\hat K$
vanishes, while maintaining the  condition $\mcK(T^3)=\{0\}$.

Now let $\tilde M$ consist of two copies of $T^3$, with initial data
$(\hat \gamma, \hat K)$ on the first copy, $T^3_1$, and with data
$(\hat\gamma, -\hat K)$ on the second copy, $T^3_2$. We let
$\Omega_a=T^3_a$ for $a=1,2$  and we let 
$p_a$ denote the points in $M_a$ corresponding to $p$.
Noting that the mean curvature vanishes in symmetric neighborhoods of
$p_1$ and $p_2$, we may now apply an argument similar to that used in 
proving Theorem~\ref{Tlgluingv} to this initial
data set on $\tilde M$ relative to the points $p_a$. For this procedure
to produce the
desired initial data set on $T^3\#T^3$, it is crucial to verify that
all the steps are done with the
correct symmetry around the middle of the connecting neck. 
In particular, we must check that the glued data
obtained from this procedure gives 
a solution of the constraints which has the symmetry indicated by the 
presence of the reflection map $\mu$.
This is ensured by using approximate solutions with the
same reflection symmetry in the construction used in the first 
step of the proof of
Theorem~\ref{Tlgluingv}. That the end result has the same symmetry 
follows from the uniqueness (within the given conformal class)
of the solutions obtained there.

\section{Conclusions}

We see from this result that gluing is a powerful tool for the 
mathematical analysis of solutions of Einstein's equations. Other results 
relying on gluing, such as the proof that for any closed manifold $\Sigma$ 
there exists an asymptotically Euclidean \cite{IMP2}
as well as an asymptotically hyperbolic solution \cite{IMP}
of the constraints on $\Sigma$ with a point removed, 
support this contention. The notion of gluing which we have explored here, 
which is topologically the connected sum or `handle addition' (for wormholes)
operation, is the simplest sort of surgery one can perform on manifolds.
In space-time dimensions $n+1$, for $n>3$, it is likely that similar 
results can be established for other types of surgeries. 
This extension would be of interest, for example, 
in the construction of black strings \cite{Hor} 
and will be discussed elsewhere.
Not yet fully explored is the extent of the  utility of this 
procedure for constructing physically interesting solutions. 
Theorem \ref{Tlgluingv} together with results from \cite{ChDelay} 
shows that we can use gluing to produce a wide variety of multi 
black hole solutions with prescribed asymptotics. We can also use it 
to glue a prescribed black hole to a cosmological solution. Will these 
glued solutions be useful for modeling astrophysical or cosmological 
phenomena? We believe so, and we are working to demonstrate this utility.

\bibliography{gluing}

\end{document}